\documentclass[iop]{emulateapj-rtx4} 
\shortauthors{Sekanina}
\shorttitle{Formation and Evolution of SOHO Kreutz Sungrazers. I}
\slugcomment{Version \today }
\newcommand{\Rsun}{$R_{\mbox{\scriptsize \boldmath $\odot$}}\!$}

\begin{document}
\title{Formation and Evolution of the Stream of SOHO Kreutz Sungrazers.\\
 I. History and Preliminary Investigations \\[-1.5cm]}
\author{Zdenek Sekanina}
\affil{La Canada Flintridge, California 91011, U.S.A.; {\sl ZdenSek@gmail.com}}

\begin{abstract} 
The nearly continuous stream of miniature comets dominated by the
Kreutz sungrazers has been an unexpected great bonanza for cometary
science initiated by the launch of the Solar and Heliospheric Observatory
(SOHO) in 1995.  Over the nearly 30~years since the time, no serious attempt
has~been made to formulate a self-consistent model for the formation and
evolution of this stream of Kreutz comets --- the goal of the present two-part
investigation.  Part~I describes historical highlights of the research
that has been relevant to the problem of SOHO sungrazers (including the major
contributions~by Hubbard, Kreutz, and Marsden) and furnishes preliminaries
of diagnostic value that are intended to facilitate, and provide critical
information for, the work in Part~II.  Formerly noted issues, such as the
high frequency of close pairs in the SOHO database, are proposed to be
products of a broader process of swarming, seen in both the nodal longitude
and time.  I present examples of tight swarms revealed by high arrival rates
of the SOHO Kreutz sungrazers, primarily from Population~I.{\vspace{-0.05cm}}

\end{abstract}

\section{The History} 
Knowledge of the Kreutz sungrazer system, a stream of comets that
approach the Sun to less than~2~{\Rsun}~\,at~peri\-helion and move in
orbits of a joint line~of~\mbox{apsides}, has grown explosively since
the beginning of the~era~of~Sun-exploring space missions.  Whereas
the number~of~the known members did not exceed a
dozen (including~a~few questionable ones) in the late 1960s,
as~noted~by~\mbox{Marsden} (1967), it grew to nearly 30 by 1990, to
more~than~350 by the end of 2000, to more than 1700 by the end
of 2010, and to more than 4300 by the end of~2023,~thanks~mainly
to imaging~with~the~C2~and~C3~coronagraphs on board
the {\it Solar and Heliospheric Observatory\/} (SOHO).  Very
recently, sungrazers undetected by SOHO were observed by the
{\it Parker Solar Probe\/} (e.g., R.\ Pickard's remark in Facebook's
{\it Sungrazer Comet Section\/} on 2023~Nov.~9).\,\,

Even though this system of sungrazing comets has been named after
him, H.~Kreutz was not the first scientist to study the orbital
properties of the members in some detail.  The motion of the Great
March Comet of 1843 (C/1843~D1) had extensively been examined by
Hubbard (1849, 1850, 1851, 1852), who accounted for the perturbations
by six planets (Mercury to Saturn) and derived two sets of orbital
elements; one based on all available astrometric observations made with
a filar micrometer, the other with either a filar or ring micrometer.
For an osculating orbital period Hubbard obtained in the two cases,
respectively, 533$\pm$135~yr (Elements~VII; the mean error converted from
the used probable error) and 803~yr (Elements~VI; with no error given,
but the uncertainty admittedly somewhat higher). 

Kreutz's (1901) ``definitive'' orbit for this 1843 sungrazer, with
the planetary perturbations unaccounted for and the observations
lumped up into only seven normal places, has been less informative
than Hubbard's.  However, a useful feature of Kreutz's paper was
the discrimination of the astrometric observations into quality
categories, exploited by Sekanina \& Chodas (2008) to compute
the comet's new set of orbital elements by using only the
best-quality data with the comparison-star positions from the
modern catalogues.  These results offered
an optimum osculating orbital period of 654$\pm$103~yr.  A
forced period of 742~yr, needed in order to successfully link
C/1843~D1 with X/1106~C1, increased a mean residual very
insignificantly, from $\pm$2$^{\prime\prime}\!$.57 to
$\pm$2$^{\prime\prime}\!$.58.  Varying a cutoff for the residuals
from 2$^{\prime\prime}$ to 7$^{\prime\prime}$ offered
nominal osculating periods between 625~yr and 790~yr.

As a witness to the arrivals of additional spectacular members
of the sungrazer system in the 1880s, Kreutz was in a position
that Hubbard (1823--1863) was denied.  And it should be emphasized
that Kreutz's (1891) orbital solutions for the four nuclear
fragments of the Great September Comet of 1882 (C/1882~R1) observed
extensively after perihelion, are a masterpiece, especially given
that they are based on mediocre observations by today's standards.
Of particular interest is his orbit for fragment~B (or No.~2),
the best candidate~by~far~for~the~principal nucleus (presumably
the most massive object of the four).  Marsden's (1967)
integration of its orbit back to the 12th century indicated that
perihelion was reached in April 1138, merely a few months (!) earlier
than the estimated perihelion time of the Chinese comet of
1138 [No.~403 in Ho's (1962) catalogue], proposed by Sekanina
\& Kracht (2022) to have been the previous appearance of
C/1882~R1 and Ikeya-Seki (C/1965~S1).

\begin{figure*}[t] 
\vspace{0.15cm} 
\hspace{-1.05cm}
\centerline{
\scalebox{0.8}{
\includegraphics{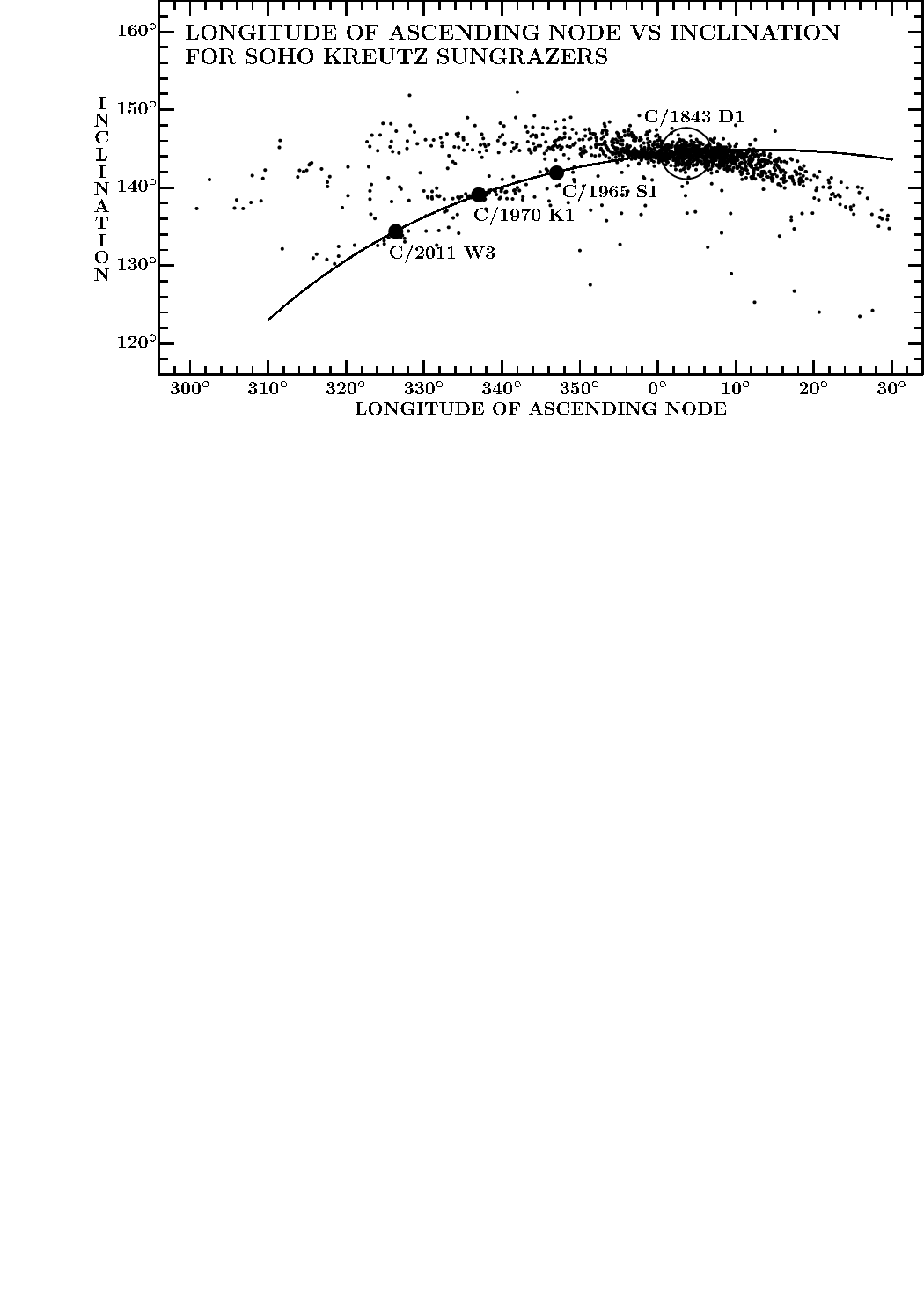}}}
\vspace{-13.85cm}
\caption{The relationship between the nominal longitude of the ascending
node and orbit inclination for more~than~1500~SOHO~Kreutz sungrazers
for which a parabolic orbit was computed by Marsden.  The Great March
Comet of 1843 is located at the center of an~oversized open circle, while
the positions of three bright sungrazers --- C/1965~S1, C/1970~K1, and
C/2011~W3 --- are marked by the large~solid circles.  The passing curve
connects all orbits whose line of apsides has the same position, the
perihelion coordinates given by \mbox{$L_\pi \simeq 283^\circ$, $B_\pi
\simeq +35 ^\circ$} (Equinox J2000). The dots mark the SOHO sungrazers.
Most of them make up a thick band that runs through the position of
C/1843~D1 and would line up along the apsidal-line curve if rotated by about
15$^\circ$ to 16$^\circ$ counterclockwise.  Sparcely populated branches
of data, parallel to the main band, are seen to pass through the positions
of the three bright sungrazers as well. (From Sekanina \& Kracht
2015.){\vspace{0.68cm}}}
\end{figure*}

Unlike many of his contemporaries, Kreutz (1901) was convinced
that the sungrazers observed between 1843 and 1887 were distinct
fragments of a common parent, not returns of the same object.
And even though he commented on the differences
in their nodal longitudes to support his view, he was not in
a position to discern the subgroups.  That could only be
accomplished after reliable orbits were determined for additional
sungrazers, which arrived in the second half of the 20th century.
It was Marsden (1967) and others who seized the oppor\-tu\-nity and
recognized the existence of Subgroups~I~and~II.

Of the sungrazers whose orbits were determined accurately or
fairly accurately, C/1843~D1, C/1880~C1, and C/1963~R1
belonged to Subgroup~I, while C/1882~R1 and C/1965~S1 to
Subgroup~II.  Unfortunately, a disconcerting pattern of
perpetually catching up with the ever widening range of
orbital properties of newly discovered naked-eye sungrazers
seems to have begun to proliferate after Marsden's (1967)
pioneering work:\ only three years later, a new bright
sungrazer, White-Ortiz-Bolelli (C/1970~K1), was discovered
to move in an orbit that defied the classification into
two subgroups, compelling Marsden (1989) to introduce
a third subgroup, IIa.  And one year after Marsden's
untimely passing away, T.~Lovejoy discovered another
sungrazer that became visible to the naked eye,
C/2011~W3, moving in an orbit that defied the expanded,
three-subgroup classification, necessitating the
introduction of a fourth subgroup, III (Sekanina \& Chodas
2012).

Even though it remains to be seen whether this perplexing
pattern will continue in the future, the chance is that
it will not.  Clear footprints of the Kreutz system's
morphology much more complex than formerly believed became
in the meantime apparent from a select subset of the SOHO
Kreutz database, which offered compelling evidence not only
for the four subgroups noted above, but for five additional
ones as well (Sekanina 2021, 2022).  At the same time, as
the Kreutz membership was exponentially climbing first into
hundreds and more recently into thousands of objects, the term
{\it subgroup\/} seemed ever less appropriate.  Eventually
I replaced it in my papers with a term {\it population\/}.

\section{The SOHO Kreutz Database and Orbital\\Properties
of the Sungrazers} 
Most Kreutz sungrazers that we know are boulders estimated
at 10~meters or less across as they are detected on their doomed
journey to perihelion.  Indeed, none has as yet been observed to
survive, all having disintegrated shortly before reaching it.
Because of their tiny dimensions, they are often referred
to as {\it dwarf\/} sungrazers.  Only a single Kreutz
comet visible to the naked eye from the ground and marginally
surviving perihelion has been observed in the 50+~years since
1970.

%
More than 20 years ago I remarked that too many close pairs of SOHO
Kreutz objects had been reported to explain their existence as random
events; instead, these instances of episodic nature appear to have been
products of secondary (or higher-generation) fragmentation at large
heliocentric distance (Sekanina 2000).  While this conclusion is hard
to argue with, the details (such {\vspace{-0.03cm}}as the separation
velocities of up to 6~m~s$^{-1}$) could be problematic.  And in any
case, most SOHO Kreutz sungrazers have not arrived at perihelion in
very close pairs, so the {\it probability\/} of occurrence of such
fragmentation episodes would appear to be low.  In the early days of
SOHO's operations, little was known about the meaning and usefulness of
Marsden's orbital elements, about the magnitude of the nongravitational
acceleration, and especially about the fragmentation history of the
Kreutz system and its populations to offer far-sighted predictions.

\begin{figure*}[t] 
\vspace{0.15cm}
\hspace{0cm}
\centerline{
\scalebox{0.8}{
\includegraphics{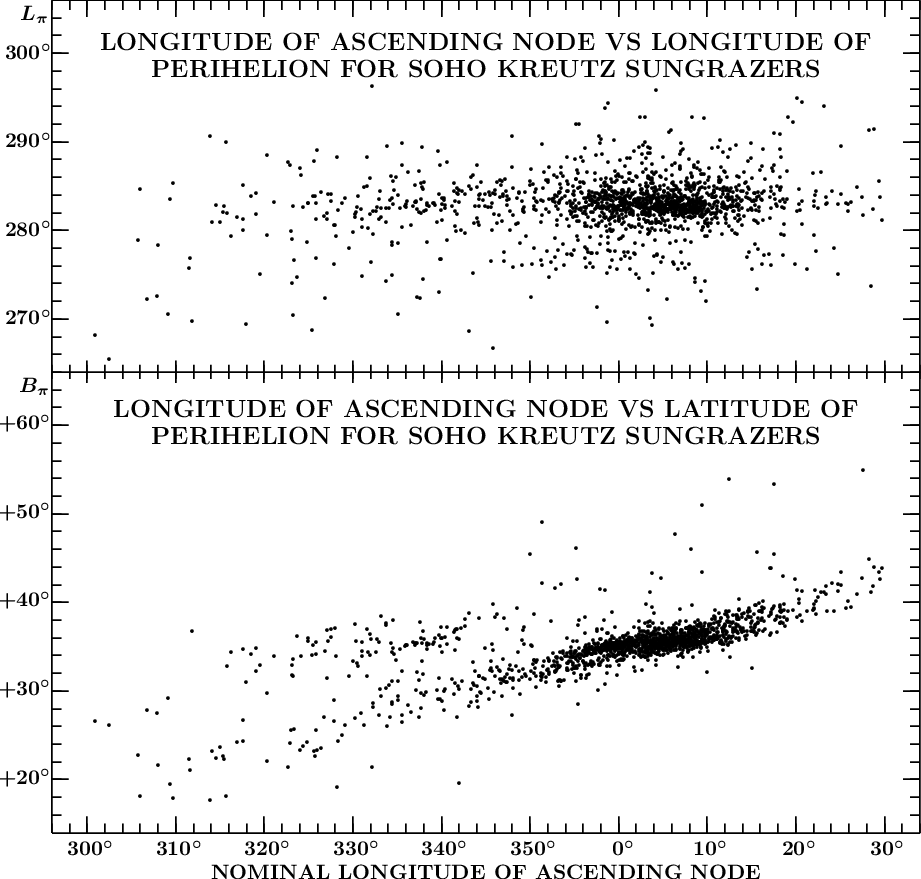}}}
\vspace{-0.3cm}
\caption{The relationships between the nominal longitude of the ascending
node, on the one hand, and the perihelion longitude,~$L_\pi$,~and
perihelion latitude, $B_\pi$, on the other, for the same 1500+ SOHO Kreutz
sungrazers.  While the perihelion longitude is seen to be statistically
independent of the longitude of the ascending node, the perihelion
latitude appears to be linearly increasing at a rate of approximately
3$^\circ$ per 10$^\circ$ increase in the longitude of the ascending
node. Note the secondary branches above the main band. (From Sekanina
\&~Kracht~2015.){\vspace{0.7cm}}}
\end{figure*}

\subsection{The Line of Apsides Paradox} 
Once Marsden's orbits were available for a large number of SOHO Kreutz
sungrazers (Marsden \& Williams 2008), a baffling pattern became apparent
in a plot of the inclination against the longitude of the ascending node
in Figure~1:\ the condition of a shared position of the line of apsides,
often serving in orbital studies as a sign of potential genetic
relationship between two or more independent objects, and satisfied
by all bright members of the Kreutz system observed from the ground,
is ostensibly defied by most SOHO sungrazers, which are seen to line
up along a curve that makes up an angle of 15$^\circ$ to 16$^\circ$
with the curve of the fixed apsidal line.  Only the SOHO sungrazers
located in the figure in close proximity of C/1843~D1 satisfy, at least
approximately, the apsidal-line condition.  This likewise applies to the
SOHO sungrazers in the figure that relate to any of the other three bright,
naked-eye objects in the same manner as do the above SOHO sungrazers
in the main band to C/1843~D1.

The paradox suggests that the least-squares fit by a set of
purely gravitational parabolic elements (referred to in this paper as
{\it nominal\/}), applied by Marsden to satisfy the measured positions
of a SOHO sungrazer over the short arc of its path, was forced to respond
to the strong orbital constraint from a culprit at the expense of the
more subdued condition of the shared line of apsides.  The nature of
the culprit is not difficult to guess, given the minuscule dimensions
and relatively high activity of the SOHO dwarf sungrazers:\
an obvious candidate is a major, sublimation-driven nongravitational
acceleration.  Furthermore, given that the paradox is involving a plot
of two {\it out-of-plane\/} elements, the culprit apparently was the
{\it neglected\/} nongravitational acceleration's normal component.
This suspicion appears to be confirmed by plots of the nominal
perihelion longitude, $L_\pi$, and latitude, $B_\pi$, against the
nominal longitude of the ascending node, $\Omega$, displayed in
Figure~2:\ whereas the peri\-helion longitude is statistically
constant, independent of the longitude of the ascending node, the
perihelion latitude varies with~the nodal longitude systematically,
in a linear fashion.

Sekanina \& Kracht (2015) selected eight individual SOHO Kreutz
objects, imaged exclusively in the C2 coronagraph, whose nominal
longitudes of the ascending node spanned the entire range in
Figures~1 and 2, from 300$^\circ$ to 30$^\circ$, at intervals
that averaged some 10$^\circ$ to 15$^\circ$.  The task was to fit
each object's motion by applying Marsden et al.'s (1973) standard
nongravitational model (Style~II), forcing the normal component of
the water-ice sublimation acceleration, $A_3$, and searching, by
trial and error, for an orbital solution offering an orientation
of the line of apsides that agreed with the standard position as
closely as possible.  This orbital solution provided representative
estimates for values of the {\it true\/} elements and for the parameter
$A_3$ of the presumably dominant out-of-plane component of the
nongravitational acceleration.

The results for the eight selected objects were consistently astonishing:\
(i)~the {\it true\/} orbital elements strongly resembled --- the
angular ones within degrees --- those of C/1843~D1, which meant that in
the extreme cases the longitude of the ascending node got changed by
50$^\circ$ or more (!) from the nominal value; (ii)~the out-of-plane
nongravitational acceleration was on the average at least three orders
of magnitude higher than the radial nongravitational accelerations in
the motions of ordinary comets and the peak value was {\it comparable
to the solar gravitational acceleration\/} at the given heliocentric
distance; (iii)~the standard position of the line of apsides was always
closely approximated; and (iv)~the quality of the fit provided by the
nongravitational orbit to the used C2 coronagraphic observations,
expressed in terms of a mean residual, was comparable to that achieved
by Marsden's nominal gravitational orbit.

The paradox was thereby fully resolved.

\subsection{Populations of SOHO Sungrazers} 
I recognized that the apparently linear relationship between the
nominal perihelion latitude and the nominal longitude of the ascending
node in Figure~2 could potentially offer a gold mine of valuable data on
the Kreutz system in the SOHO database.  The weak point was considerable
data scatter involved.  Fortunately, much of it could be removed by
limiting the acceptance to the SOHO sungrazers imaged exclusively by the
C2 corona\-graph, whose sensor has a pixel size nearly five times smaller
(linearly) than the C3 coronagraph, thereby providing the much better
astrometry.  The apparent orbital arc in C2 is shorter, but only because
of projection effects. On the other hand, the terminal part of the orbit
is the one with the greatest curvature.  The major shortcoming is
the missing data from the months of January through March and July
through September, when the path of a Kreutz comet in space is nearly
perpendicular to the SOHO's line of sight and the dwarf sungrazers
disintegrate before they enter the C2 coronagraph's field of view.  To
avoid unwanted effects of this kind, one should not get involved with
distributions whose sampled base is divided into intervals shorter
than one year but longer than several weeks.

\begin{table}[b] 
\vspace{0.6cm}
\hspace{-0.2cm}
\centerline{
\scalebox{1}{
\includegraphics{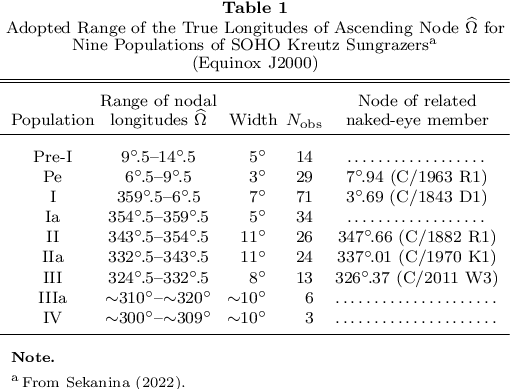}}}
\vspace{-0.08cm}
\end{table}

Examining the relationship between the nominal longitude of the ascending
node, $\Omega$, and perihelion latitude, $B_\pi$, among the SOHO
sungrazers imaged exclusively in the C2 coronagraph, I was surprised
to see how much did the morphology of the thick band in Figure~2~change.
The data looked no longer smeared as they did before the objects whose
orbits were based on the C3 coronagraphic images were eliminated.
Significantly, it now became obvious that the data did not refer
to a single~\mbox{population}.  On the contrary, the band of the data
appeared to be resolved into {\it four\/} discrete populations, which
I designated I, Ia, Pe, and~\mbox{Pre-I}.  The boundaries between
them became still clearer when the plotted entries were
restricted to the objects with better orbits, based on a minimum
of 12~observations for Populations~I and Pe, 11~observations for
Population~\mbox{Pre-I}, and 10~observations for Population~Ia.  For
the remaining five populations, all objects imaged at least five times
were included.  The complete set of the nine populations totaled 220
select SOHO Kreutz sungrazers.  The memberships in Populations~II, IIa,
III, IIIa, and IV, were gradually declining (\mbox{Sekanina} 2021, 2022).
Populations~I, II, and IIa were identical with Marsden's (1967, 1989)
Subgroups~I, II, and IIa, respectively.  The total number of the SOHO
Kreutz sungrazers observed exclusively in the C2 coronagraph between 1996
and mid-2010, for which a parabolic orbit was determined by Marsden,
was 753.  Most~of~them, a little over 50~percent, belonged to Population~I.

In the $B_\pi(\Omega)$ plot, the members of the nine populations were
distributed along parallel straight lines given by a common slope
of \mbox{$dB_\pi/d\Omega = 0.28$}, which equals tan~15$^\circ\!$.6,
the angle between the two curves in Figure~1 (Sekanina 2021).  For
each SOHO object a {\it true\/} longitude of the ascending node was
calculated from its nominal value, the slope, and a standard value
of the perihelion latitude.  I assigned each population a range of
the true longitudes of the ascending node (Table~1) and used the
220~objects to construct a histogram of the populations (Figure~3),
a major attribute of the Kreutz system among the SOHO sungrazers.

\subsection{Implications} 
The investigation of the eight selected SOHO objects by Sekanina \&
Kracht (2015) and the subsequent examination of the $B_\pi(\Omega)$
plot in Figure~2 thus showed that the longitude of the ascending
node is a critical orbital element that describes the population and
provides information on the fragmentation process and the evolution
of the Kreutz system.  This conclusion was not obvious after the
first two or even three subgroups were defined by Marsden (1967,
1989), because the nodal line correlated with the perihelion
distance.  It was only the introduction of Population~III
that showed that the perihelion distance was not
diagnostic, as there was little difference in this element
between Comet Lovejoy (C/2011~W3; Population~III) and the Great
March Comet of 1843 (C/1843~D1; Population~I).

\begin{figure*}[t] 
\vspace{0.15cm}
\hspace{-0.5cm}
\centerline{
\scalebox{0.8}{
\includegraphics{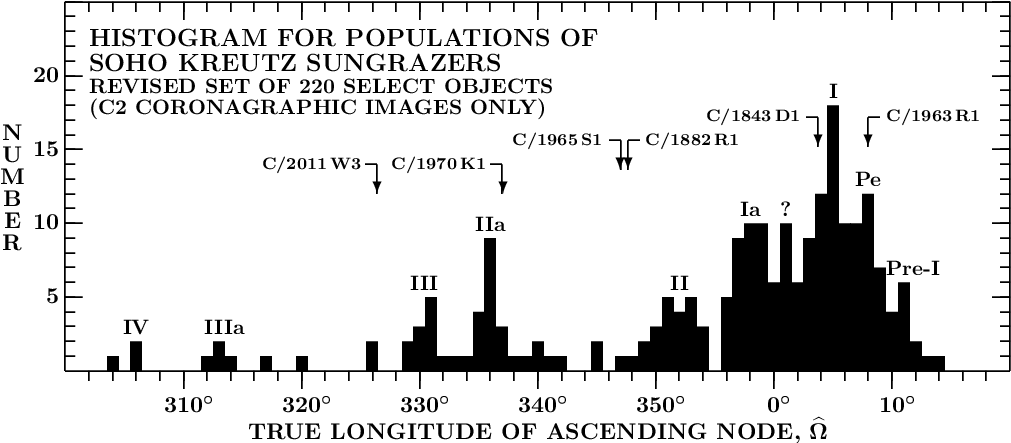}}}
\vspace{-0.1cm}
\caption{Histogram of the true longitudes of the ascending node of 220
select SOHO sungrazers, imaged exclusively by the C2 coronagraph,
showing nine populations of the Kreutz system.  A peak at 1$^\circ$,
marked by a question mark, is probably a side branch of Population~I,
with no known associated naked-eye object.  (From Sekanina
2022.){\vspace{0.8cm}}}
\end{figure*}

The {\it true\/} perihelion distances of most SOHO Kreutz sungrazers
are not known at this time, because no procedure is available for
converting the nominal perihelion distances.  Sekanina \& Kracht
(2015) indicated no systematic corrections to these values as a
function of the nodal longitude.  Thus, the typical perihelion distances
of the populations whose major sungrazers have never been seen ---
\mbox{Pre-I}, Ia, IIIa, and IV --- remain unavailable.  And for
Population~IIa, for which C/1970~K1 suggests a perihelion distance
of 1.91~{\Rsun}\,, the perihelion-distance distribution of the SOHO
sungrazers appears to potentially suggest two different sources, the
other, IIa$^\star$, having a perihelion distance similar to that of
C/1843~D1.

Overall, the agreement in the position of the nodal lines (and, by
definition, in the values of the other two angular elements) between
a bright Kreutz sungrazer (or sungrazers) and the related SOHO objects
of the same population is very good, as is apparent from Table~1 and
Figure~3.

To finish this section, a few words about the limitations in the use of
Marsden's parabolic orbits for ephemerides of the SOHO sungrazers and
similar purposes.  As long as the nodal longitude and orbit inclination
show that the object's position in Figure~1 is very close to the
apsidal-line's curve, Marsden's elements are likely to provide a fair
approximation to the actual path in space.  If this condition is not
satisfied, the elements should not be used for such purposes.  In these
cases one can only trust the {\it true\/} longitude of the ascending node
converted from the nominal value (with help of the standard coordinates
of the line of apsides) as explained.  In addition, the nominal perihelion
time could approximately be converted into a true perihelion time by
applying the following correction (in days), derived from the eight
investigated objects by Sekanina \& Kracht (2015):
\begin{equation}
{\sf corr}(t_\pi) = -0.0037 - 0.0017 (\Omega \!-\! \widehat{\Omega}),
\end{equation}
where $\Omega$ and $\widehat{\Omega}$ are, respectively, the nominal
and true longitudes of the ascending node{\vspace{-0.07cm}} (in
degrees).  The correction applies in the range of \mbox{$-70^{\circ}\!
< \Omega \!-\! \widehat{\Omega} < \!+20^\circ$} and it is expected to
be good to $\pm$0.01~day.

\section{Temporal and Spatial Distributions of\\SOHO Sungrazers} 
Battams \& Knight (2017) presented the statistics of the {\it Sungrazer
Project\/} up to the end of 2015.  Because they did not distinguish
between the Kreutz sungrazers seen in C2, C3, and combined C2+C3,
and ignored the populations, their results are mostly irrelevant to
the present investigation, except for the numbers of full-resolution
images returned per year separately for the C2 and C3 coronagraphs.
Their Table~3 shows that among the comets for which Marsden computed
the orbits (1996 through mid-2010), a meaningful statistical sample
should be based on the data from the period of 2000--2009, when the
number of useful images returned by C2 was in a fairly narrow range
between 21.2 and 24.5 thousand per year.

I determined the number of Kreutz sungrazers in each of the nine
populations detected each year between 2000 and 2009, and scaled
these numbers to 25~thousand useful images per year. I thereby
obtained the normalized statistics of SOHO Kreutz discoveries
displyed in Figures~4 and 5.  One does not need to apply any
rigorous statistical criteria to readily discover enormous
variations from year to year in any of the populations whose
total number of members in the set is large enough that such
variations should be deemed meaningful.  Three populations with
a total of about 20~members or more --- \mbox{Pre-I}, II, and
IIa --- suggest that no members were observed in some of the
years, while a peak rate of each population in a single year
equaled $\frac{1}{4}$ its total number of objects observed in the
entire decade.  For Population~Ia, growing rapidly with time, and
Pe the maximum-to-minimum rate ratios were nearly 10:1 and 4:1,
respectively.

\begin{figure}[t] 
\vspace{0.2cm}
\hspace{-0.18cm}
\centerline{
\scalebox{0.698}{
\includegraphics{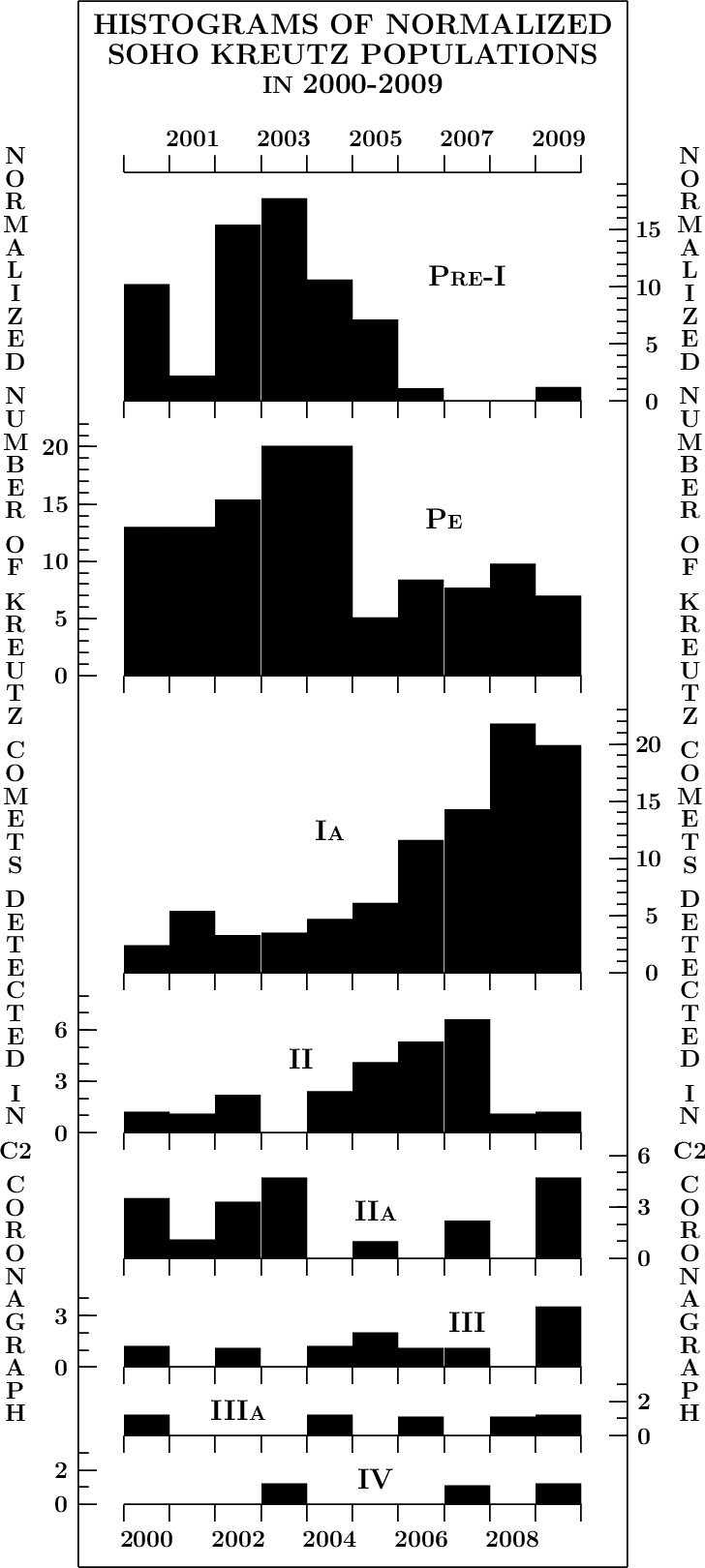}}}
\vspace{0.4cm}
\caption{Histograms for eight populations of SOHO Kreutz sungrazers
 from the years 2000--2009, imaged exclusively in the C2 coronagraph
 and for which a parabolic orbit was computed by Marsden.  The data
 are normalized to an equal number of useful images per year.  Note
 the large, uneven year-to-year variations.{\vspace{0.4cm}}}
\end{figure}

For Population I, the most abundant one among the SOHO sungrazers,
the maximum-to-minimum rate ratio over the period of 10~years was
2.3:1, peaking in 2006 on a gradually increasing background.  Among
the members are eight 2006 objects (C/2006~K18, K19, K20, L3, L4, L6,
M6, and M9) that arrived at perihelion between May~26 and June~27
(see Table~5 of Sekanina 2022).  The 56$^\circ$ wide interval of the
nominal nodal longitudes (from Marsden's parabolic orbits), ranging
from 345$^\circ$ to 41$^\circ$, was reduced --- after the effects
of the out-of-plane component of the nongravitational acceleration
had been allowed for --- to merely 6$^\circ$, centered on a true
nodal longitude of 3$^\circ$, a fraction of a degree away from
the nodal longitude of the Great March Comet of 1843.

Figure 5 compares the histogram for the genuine Population~I
with the histogram of the combined Populations~I, Ia, Pe, and
\mbox{Pre-I} --- what could formally be called Population~I+, based
on the appearance in the plot that includes the low-quality C3
data.  Remarkably, even though each of the four populations that
make up Population~I+ displayed profound variations in the C2 data
between 2000 and 2009, their sum was much flatter, exhibiting a
maximum-to-minimum rate ratio of only 1.5:1 and in the years 2003,
2004, 2006, 2008, and 2009 their combined rates being nearly constant.

\begin{figure}[t] 
\vspace{0.2cm}
\hspace{-0.2cm}
\centerline{
\scalebox{0.694}{
\includegraphics{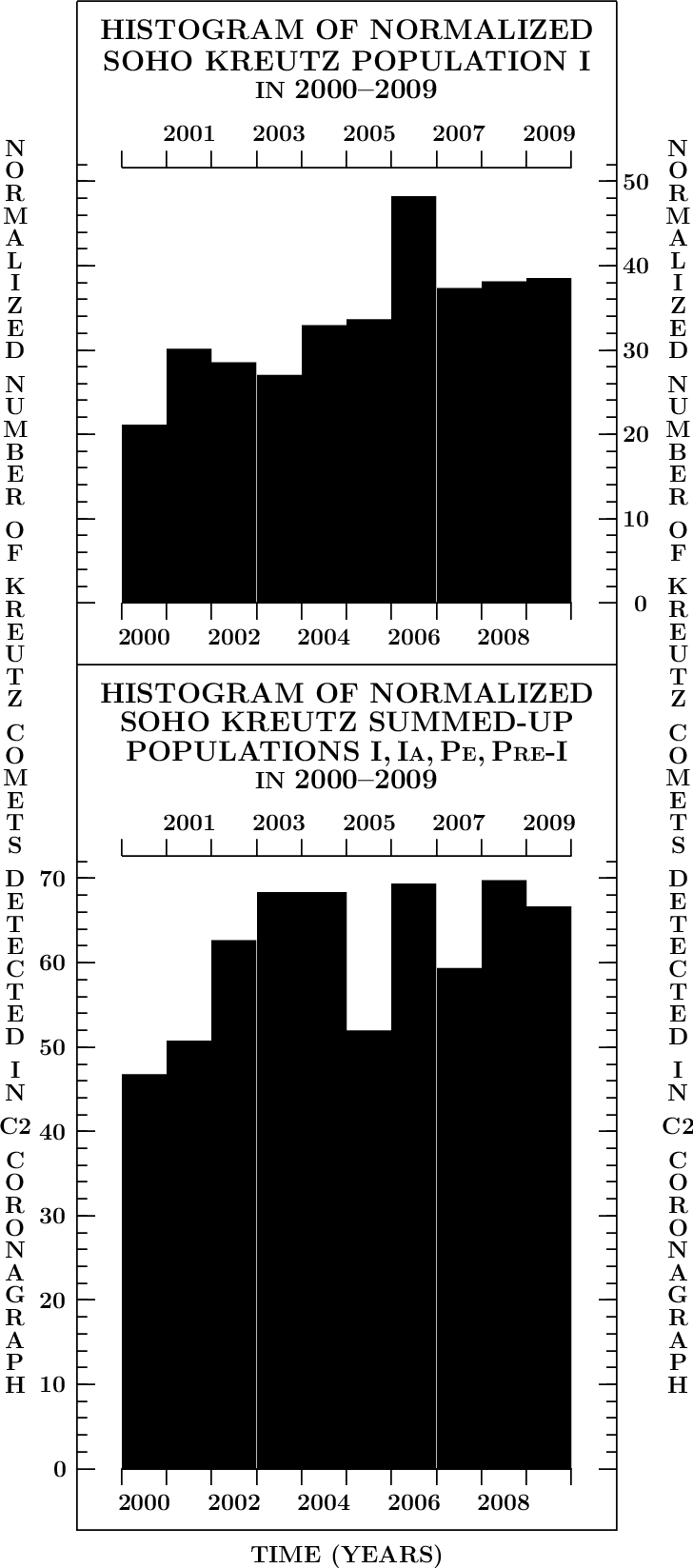}}} 
\vspace{-0.1cm}
\caption{Histogram for Population I (top) and a sum of Populations~I,
 Ia, Pe, and Pre-I of SOHO Kreutz sungrazers from the years 2000--2009,
 imaged exclusively in the C2 coronagraph and for which a parabolic
 orbit was computed by Marsden.  The data are normalized to an equal
 number of useful images per year.{\vspace{0.45cm}}}
\end{figure}

To sum up the findings from Figures 4 and 5, the histograms display
very strong relative variations in the incoming flux of the individual
populations of SOHO Kreutz sungrazers on a time scale of years, perhaps
even somewhat shorter.  The interpretation of this important result is
left for Part~II.  Here I focus on collecting all evidence that supports
\begin{figure*}[t] 
\vspace{0.2cm}
\hspace{-0.2cm}
\centerline{
\scalebox{0.76}{
\includegraphics{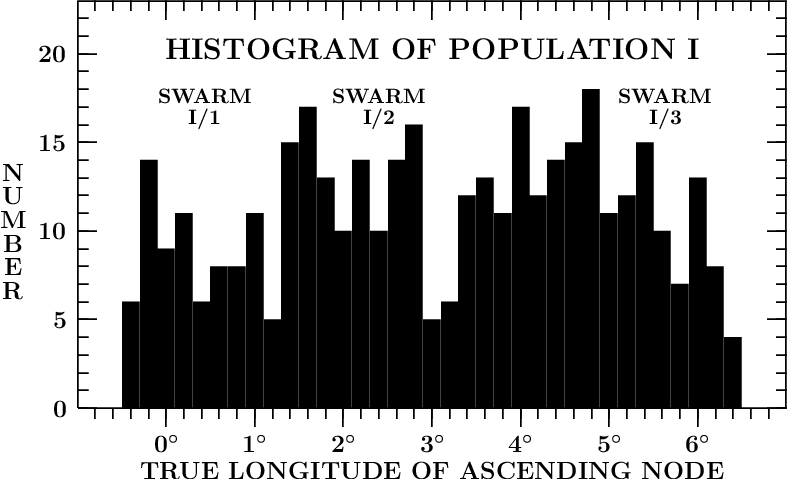}}}
\vspace{-0.1cm}
\caption{Histogram of true longitudes of the ascending node for
390 SOHO Kreutz sungrazers of Population~I, imaged exclusively in the
C2 coronagraph from 1996 to mid-2010.  Note the three major peaks,
below 1$^\circ$, between 1$^\circ$ and 3$^\circ$, and beyond
3$^\circ$, separated~by~deep~minima.{\vspace{0.6cm}}}
\end{figure*}
a self-consistent model for the formation and evolution of the stream
of SOHO Kreutz sungrazers.

Before I investigate apparent swarming\footnote{I deliberately am
avoiding the term {\it clustering\/}, which has been used in connection
with the sungrazers for another phenomenon.} of the SOHO sungrazers over
very short periods of time, a topic that essentially expands the issue
of close pairs (Section~2), I examine the phenomenon of swarming of the
sungrazers in the longitude of the ascending node.
 
\subsection{Swarming of SOHO Kreutz Sungrazers\\in the Nodal Longitude}
%
As will become obvious from Part II of this investigation, the SOHO
Kreutz sungrazers' swarming in the parametric space of the true longitude
of the ascending node (or another angular element) is as important
for understanding the formation of the sungrazers' stream as is their
swarming in time.  For the largest Population~I, a histogram of the
distribution of true nodal longitudes is displayed in Figure~6.
It includes 390~objects imaged exclusively in the C2 coronagraph
between 1996 and mid-2010, for which Marsden's nominal elements
were available.

The histogram shows that the distribution potentially consist of three
distinct swarms, centered at the nodal longitudes of, respectively, less
than 1$^\circ$,  near 2$^\circ$, and well beyond 3$^\circ$.  The last swarm
appears to be the widest and most massive.  It is likely that at least
most objects in each of the three groups are more closely
related to one another than the objects across the boundaries.  It is
impossible to follow the histories of the individual fragments but
below I use a {\it dispersion\/}, defined as a mean quadratic
difference between all fragments in a swarm, as a measure of random
scatter in the given quantity.  The dispersion{\vspace{-0.07cm}}
in the true longitude of the ascending node, {\sf disp}$(\widehat{\Omega}$),
is defined by
\begin{equation}
{\sf disp}(\widehat{\Omega}) = \sqrt{\frac{2}{n(n \!-\! 1)} \sum_{i \neq j}
 \left( \widehat{\Omega}_j \!-\! \widehat{\Omega}_i \right)^{\!2}},
\end{equation}
where $n$ is the number of objects in the{\vspace{-0.05cm}} swarm
and \mbox{$i \neq j$} emphasizes that $\widehat{\Omega}_i$ and
$\widehat{\Omega}_j$ are any two different objects in the set,
although of course in practice the result is the same regardless of
whether the rule is honored or not.

The true perihelion time, $\widehat{t}_\pi$, is approximated by a formula
\begin{equation}
\widehat{t}_\pi = t_\pi + {\sf corr}(t_\pi),
\end{equation}
where ${\sf corr}(t_\pi)$ is given by Equation~(1).{\vspace{-0.07cm}}
The dispersion in the true perihelion time, ${\sf disp}(\widehat{t}_\pi)$,
referred to from now on as the dispersion in time, equals
\begin{equation}
{\sf disp}(\widehat{t}_\pi) = \sqrt{\frac{2}{n (n \!-\! 1)} \sum_{i \neq j}
 \left[ (\widehat{t}_\pi)_j \!-\! (\widehat{t}_\pi)_i \right]^2}.
\end{equation}
I leave out a similar expression for ${\sf disp}(\widehat{q})$, because
the true perihelion distances $\widehat{q}_i$, $\widehat{q}_j$ are
unknown.

\begin{figure}[b] 
\vspace{0.65cm}
\hspace{-0.25cm}
\centerline{
\scalebox{0.83}{
\includegraphics{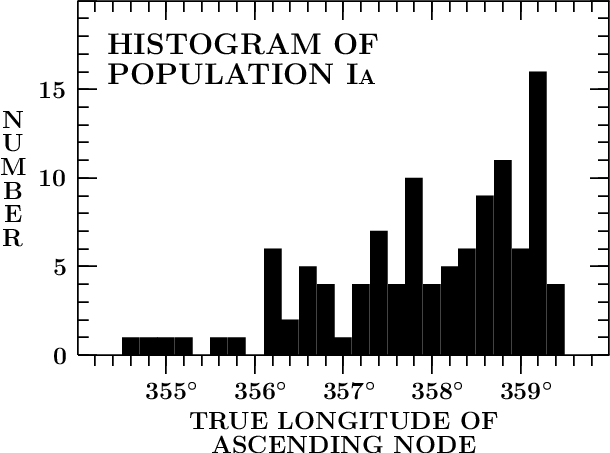}}}
\vspace{-0.05cm}
\put(101,26){\makebox(0,0){\rule{3mm}{12mm}}}
\caption{Histogram of true longitudes of the ascending node for
110 SOHO Kreutz Population Ia sungrazers, imaged exclusively in the
C2 coronagraph from 1996 to mid-2010.  Note the dramatic difference
in comparison with Population~I.{\vspace{-0.1cm}}}
\end{figure}
\begin{figure}[t] 
\vspace{0.13cm}
\hspace{-0.2cm}
\centerline{
\scalebox{0.83}{
\includegraphics{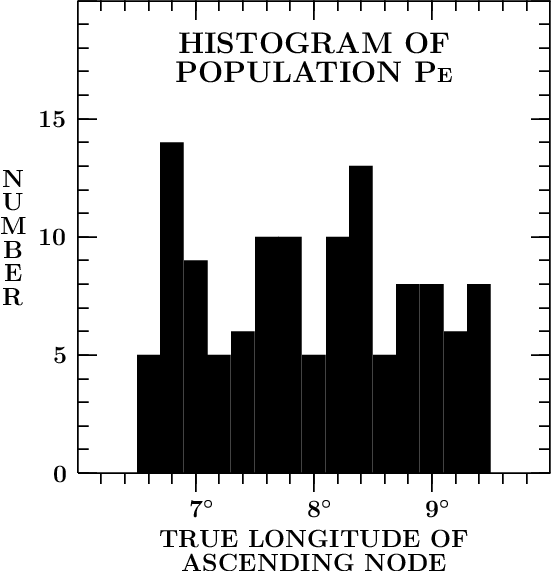}}}
\vspace{-0.05cm}
\caption{Histogram of true longitudes of the ascending node for
122 SOHO Kreutz Population Pe sungrazers, imaged exclusively in the
C2 coronagraph from 1996 to mid-2010.  Unlike the histograms in
Figures~6 and 7, the distribution is essentially random, possibly
with a very gradual decreasing tendency toward the greater
longitudes.{\vspace{0.7cm}}}
\end{figure}

It is clearly of interest to learn whether the other populations do behave
alike or differently from Population~I and what any potential disparities
may mean in terms of the populations' formation and evolution.  A meaningful
experiment of this nature could be performed only for populations with a
sufficiently large membership.  Below I have selected a minimum of
50~members, which allows comparisons of Population~I with Ia in Figure~7,
Pe in Figure~8, and \mbox{Pre-I} in Figure~9.  It turns out~that~in~terms
of the true nodal longitude, no two populations have been behaving alike.
For Populations~Ia and Pe it is impossible to discern any potential sub-swarms,
but the membership distributions in the two populations are dramatically
different:\ the number of members in Population~Ia is growing rapidly with
the nodal longitude, in the direction of Population~I.  On the other hand,
the membership distribution in Population Pe appears to  be almost entirely
random.  Population~\mbox{Pre-I} is very unusual.  It seems to consist of
two nodal-longitude streams of vastly different properties, as described
below.

\begin{figure}[b] 
\vspace{0.8cm}
\hspace{-0.21cm}
\centerline{
\scalebox{0.83}{
\includegraphics{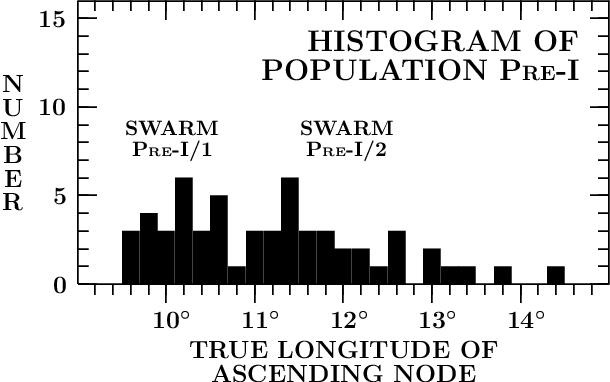}}}
\vspace{-0.05cm}
\caption{Histogram of true longitudes of the ascending node for 57
 SOHO Kreutz Population \mbox{Pre-I} sungrazers, imaged exclusively in
 the C2 coronagraph from 1996 to mid-2010.  The distribution appears to
 discriminate into two streams with contradictory properties (see the
 text).{\vspace{-0.08cm}}}
\end{figure}

The derived values of the{\vspace{-0.06cm}} dispersions ${\sf
disp}(\widehat{\Omega})$ and ${\sf disp}(\widehat{t}_\pi)$ for
the four populations and their swarms are presented in Table~2.
The dispersions in the nodal longitude for the swarms of Population~I
are approximately proportional to the chosen ranges of the longitudes,
while the dispersions in time differ insignificantly from one
another, suggesting that the sungrazers in each swarm are
distributed more or less randomly.  The dispersions in the nodal
longitude for Populations Ia and Pe are comparable given
that Pe is more compact, but the dispersions in time show
Population~Pe to be much tighter than Ia.  As already remarked,
the results for the two swarms of Population~\mbox{Pre-I} are
most controversial, with the dispersions in the nodal longitude
and time displaying sharply opposite tendencies.  The dispersion
in time for Swarm~\mbox{Pre-I/2} is particularly
anomalous, amounting to only one-third the value for
Population~Ia.  This anomaly is presumably linked to the
histogram of Population~\mbox{Pre-I} in Figure~4.  To the extent
that the swarms of a population play an important role in the
evolution of the Kreutz system, the dispersion values of that
population as a whole may become less diagnostic; two such potential
instances in Table~2 have been parenthesized.  The
significance and implications of disparities among the computed
dispersions of the populations and their swarms are left to be
discussed in Part~II.

The last numerical exercise in this subsection involves potential
detection of swarms in narrow intervals of the nodal longitude with
the highest rates of SOHO sungrazers from Population~I.  The
widths of the nodal-longitude intervals have been chosen to equal
0$^\circ\!$.2, 0$^\circ\!$.1, and 0$^\circ\!$.05 and the peak
arrival rates found to be 20, 12, and 8~objects, respectively,
for more than one interval in each group.{\vspace{-0.06cm}}
Somewhat~surprisingly, the computed dispersions ${\sf
disp}(\widehat{\Omega})$ and ${\sf disp}(\widehat{t}_\pi)$ in
Table~3 appear to be inversely correlated, with the{\vspace{-0.08cm}}
notable~exception of the interval of $\widehat{\Omega}$ between
359$^\circ\!.$83 and 359$^\circ\!.$88.\\[-0.55cm]

\begin{table}[b] 
\vspace{0.6cm}
\hspace{-0.2cm}
\centerline{
\scalebox{0.975}{
\includegraphics{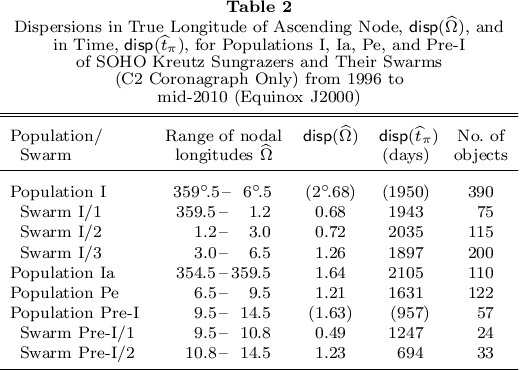}}
\vspace{0.1cm}}
\end{table}

\subsection{Swarming of SOHO Kreutz Sungrazers in Time} 
The focus of interest is next shifted toward the SOHO
sungrazers that did arrive at perihelion nearly simultaneously,
so that{\vspace{-0.075cm}} their dispersion in time is
extremely small.  I compute ${\sf disp}(\widehat{t}_\pi)$ and
${\sf disp}(\widehat{\Omega})$ for several swarms (with a minimum of four
objects), restricting this part of the study to Populations~I and
Pe only.

\begin{table}[t] 
\vspace{0.15cm}
\hspace{-0.22cm}
\centerline{
\scalebox{1}{
\includegraphics{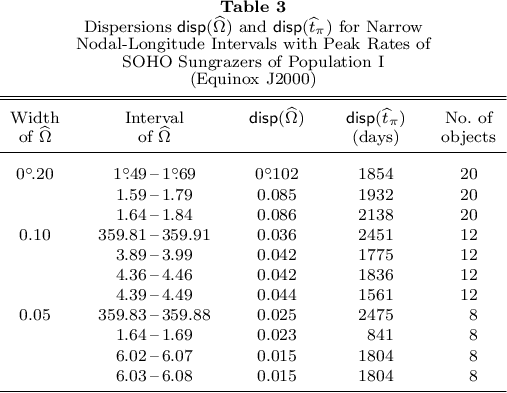}}}
\vspace{0.6cm}
\end{table}

Also, two novelties are being introduced, the purpose of which
becomes obvious in Part~II.  One is adding the dispersion
in the {\it nominal}\,\footnote{The true perihelion distance
being of course unavailable.} perihelion distance, ${\sf disp}(q)$,
in units of the solar radii \mbox{(1$\!$ {\Rsun} = 0.00465~AU)}:
\begin{equation}
{\sf disp}(q) = \sqrt{\frac{2}{n (n \!-\! 1)} \sum_{i \neq j} (q_j
 \!-\! q_i)^2}.
\end{equation}
The other is inclusion in tabular material of a limited number
of the SOHO Kreutz sungrazers imaged by the C3 coronagraph,
or, rather, by both the C2 and C3 coronagraphs.  To distinguish
them clearly from the objects imaged exclusively by the C2
coronagraph, they are marked with an asterisk.

\begin{table} 
\vspace{0.15cm}
\hspace{-0.17cm}
\centerline{
\scalebox{1.014}{
\includegraphics{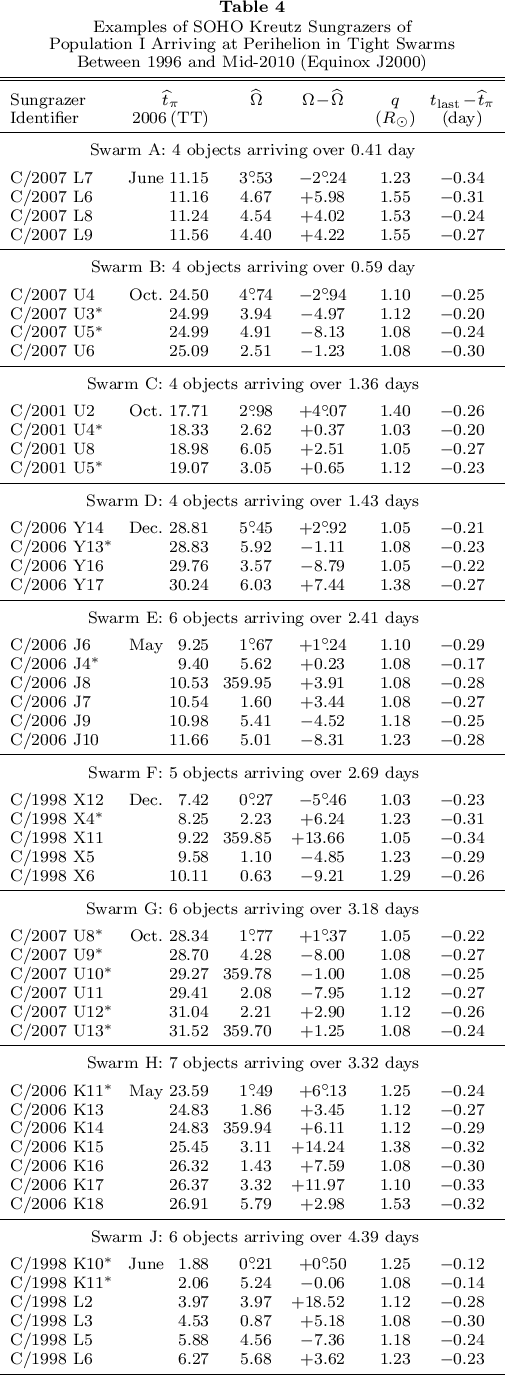}}}
\vspace{0.7cm}
\end{table}

The tabulated data would be merely a major extension of the list
of 15 comet pairs provided in Sekanina (2000), if in the meantime
the existence of the enormous nongravitational accelerations
affecting the motions of the SOHO Kreutz sungrazers
(Sekanina \& Kracht 2015) were not recognized and the method of
transforming Marsden's nominal nodal longitudes (Sekanina
2021) not developed and implemented.  Because of this progress,
a more meaningful interpretation of the SOHO database has become
possible.

The problem that cannot be completely removed from the database
is the difficulty of discriminating between objects that arrive
nearly simultaneously because until recently they were one body
or because of a coincidence.  The only modus operandi for improving
the degree of this discrimination is to incorporate the condition
of shared population.  Even this stipulation is not foolproof
because neighboring populations may partly overlap.

\begin{table}[t] 
\vspace{0.1cm}
\hspace{-0.21cm}
\centerline{
\scalebox{1}{
\includegraphics{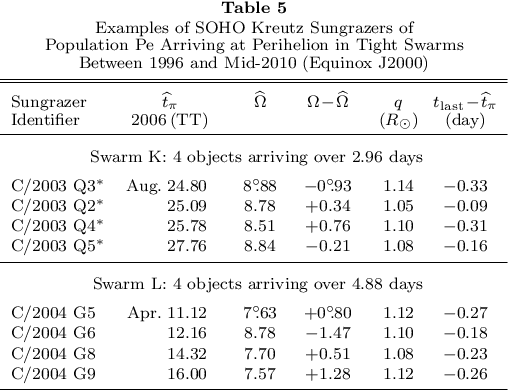}}}
\vspace{0.55cm}
\end{table}

Nine examples of SOHO sungrazers of Population~I arriving at
perihelion in tight swarms are listed in Table~4.  The individual
columns show: (1)~the sungrazer's oficial identifier (with an asterisk
when the object's orbit is not based exclusively on the astrometry
from images in the C2 coronagraph); (2)~the true perihelion time that
generally differs slightly from the value in Marsden's gravitational
orbit, being derived from Equation~(3) using the correction from
Equation~(1); (3)~the true longitude of the ascending node; (4)~the
difference between the nominal and true nodal longitudes; (5)~the
uncorrected perihelion distance (in units of the solar radius); and
(6)~the time gap between the last image in C2 and peri\-helion.

The listed swarms show interesting features.  The tightest Swarm~A
includes four sungrazers that would have passed perihelion over a
period of less than 10~hours, two within 15~minutes of each other!
The true longitudes of the ascending node of the four objects span
a range only slightly wider than 1$^\circ$, all belonging to Swarm~I/3
in Figure~6.  This contrasts with a range of more than 9$^\circ$ in
the nominal nodal longitude, a huge difference.  In fact, from the
nominal values one would classify only one object to belong to
Population~I, two would be members of Population~Pe, and one of
Population~\mbox{Pre-I}.  The four sungrazers disintegrated between
about 5 and 9~hours before perihelion.  These times are typical for
nearly all SOHO Kreutz comets, only the very bright ones among them
did survive longer, to \mbox{3--4 hours} before perihelion.

The second tightest Swarm B in Table~4 spans approximately 14~hours and
two of the four sungrazers would have passed perihelion essentially
simultaneously.  Very close pairs (about 1~hour or less apart) are
seen to be part of five among the tabulated swarms, besides A and B
also D, E, and H; the last, with seven members over 80~hours, includes
two very close pairs.  Deliberately included is Swarm~J, which contains
the well-known pair of bright sungrazers C/1998~K10 and K11.

Table 5, which has the same format as Table~4, lists two examples
of tight swarms of SOHO sungrazers of Population~Pe.  They exhibit
similar properties as the tight swarms of Population~I.  I have
also noticed a swarm of three SOHO sungrazers of Population~Ia ---
C/2004~M2, M5, and M6 --- over a period of 2.2~days.  There is no
doubt that swarms could be found in any Kreutz population of SOHO
sungrazers, once a large enough set of these objects is available.

\begin{table} 
\vspace{0.1cm}
\hspace{-0.21cm}
\centerline{
\scalebox{1}{
\includegraphics{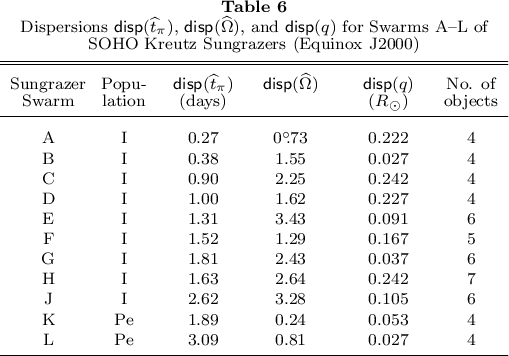}}}
\vspace{0.6cm}
\end{table}

Table 6 presents the dispersions, computed from Equations~(2), (4), and
(5), for the SOHO sungrazers' swarms listed in Tables~4~and~5.  The
dispersions in time and in the nodal longitude are very different from
those for the swarms in Table~3.  Not only does now the dispersion in
time drop three orders of magnitude and the dispersion in the nodal
longitude baloons, but their {\it trends\/} have fundamentally changed.
Whereas ${\sf disp}(\widehat{\Omega})$ and ${\sf disp}(\widehat{t}_\pi)$
for the swarms in Table~3 vary inversely with one another (separately
in each width group and with one exception), their variations for the
swarms in Table~6 are typically parallel, as shown in Figure~10.
Contrary to this remarkable peculiarity, awaiting explanation in Part~II,
the dispersion ${\sf disp}(q)$ is seen to jump up and down about
one order of magnitude, illustrating the lack of diagnostic value
of an uncorrected nominal orbital element. 

\begin{figure*}[t] 
\vspace{0.1cm}
\hspace{-0.2cm}
\centerline{
\scalebox{0.84}{
\includegraphics{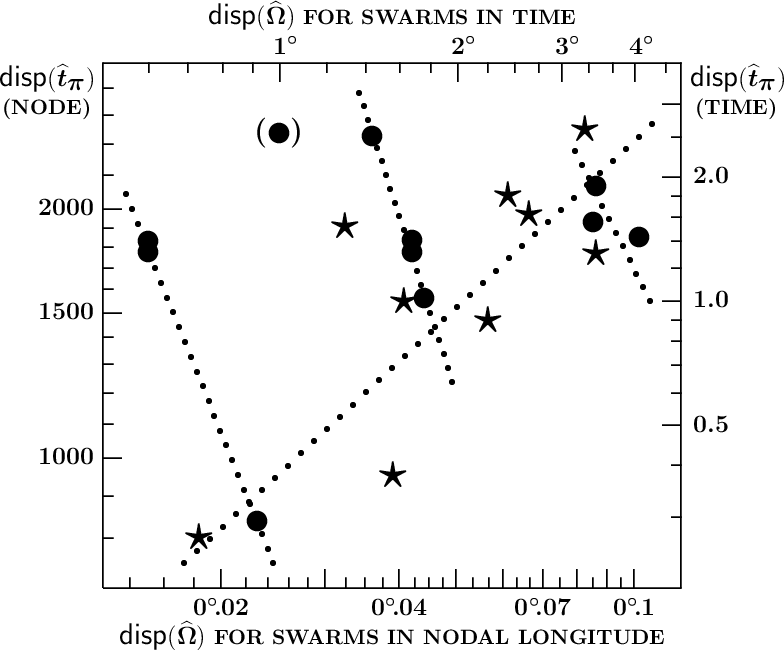}}}
\vspace{-0.11cm}
\caption{Relationships between the dispersions of swarms of SOHO
 Kreutz sungrazers of Population I in the nodal longitude and in
 time.  The circles are the swarms detected in the narrow intervals
 of{\vspace{-0.06cm}} the nodal longitude, the stars are the swarms
 in which the sungrazers arrive at perihelion almost simultaneously.
 The relation between the dispersion in the true{\vspace{-0.06cm}}
 nodal longitude, ${\sf disp}(\widehat{\Omega})$, and the dispersion
 in time (i.e., in the perihelion time), ${\sf disp}(\widehat{t}_\pi)$,
 for the two kinds of swarms is very different:\ while the dispersion
 in time decreases~with~increasing dispersion in the nodal longitude
 for the former swarms, both dispersions exhibit the same trend for the
 latter swarms.  The trends are marked with the dotted lines.  The single
 data point inconsistent with the rest is parenthesized{\vspace{0.7cm}}}
\end{figure*}

\section{Summary and Conclusions} 
Although sungrazers have always attracted much attention of astronomers,
information that actually would turn out to be relevant to in-depth
investigations of the enormous SOHO sungrazer database has historically
been accumulating very gradually.  In the mid-1800s, Hubbard was the
first to show that the Great March Comet of 1843 (C/1843~D1), which
happens to be directly associated with the bulk of the stream of the
SOHO dwarf comets, had an orbital period in a range of \mbox{500--800
yr}.  This result ruled out the then-popular hypothesis that this
object was a return of the sungrazer C/1668~E1.

Nearly a half of a century later, the understanding of the SOHO sungrazer
stream was aided by Kreutz's study of the multiple nucleus of the Great
September Comet of 1882 (C/1882~R1)\footnote{In retrospect, it is interesting
to read the section {\it On the Cause of the Partition of a Cometary Nucleus\/}
in Part~II of his monumental treatise (Kreutz 1891).  From the outset,
he kept referring to the cause as a {\it disturbing force\/} but was vague
about its nature.  Only later in the section it became clear that his
thinking was influenced by J.\ F.\ Encke's hypothesis of resisting
medium.  However, based on his computations he rejected this mechanism
and concluded instead that the disturbing force ``{\it was to be understood
as a force that developed in the interior of the comet upon its approach
to the Sun, which must have affected the individual cometary particles
differently, thereby bringing about the partition of the nucleus.\/}''
At the end of the section Kreutz admitted that ``{\it based on current
knowledge of the physical constitution of comets\/},'' the question of
the nature of the disturbing force could not be answered.} and by his
conclusion that the various spectacular 19th century sungrazers could
not be returns of the same object.  Although both results clearly implied
{\it fragmentation\/} of a cometary nucleus, Kreutz never used the term.

In the second half of the 20th century, Marsden (1967, 1989) initiated
another phase of research of what became known as the Kreutz group.
Marsden's contribution with a highly beneficial impact on studies of
the SOHO sungrazing stream was his introduction of two and later three
subgroups, for which he proposed a rationale.  Marsden also emphasized
the fixed line of apsides, shared by the members known at the time.  In
the first of his two studies he demonstrated convincingly that C/1882~R1
and C/1965~S1 (Ikeya-Seki) were one object on the way to their previous
perihelion in the 12th century, thus pointing to the propensity to
perihelion fragmentation of C/1882~R1, C/1965~S1, {\it and\/} their
parent.  In the second paper, Marsden skillfully incorporated the early
Solwind and Solar Maximum Mission (SMM) coronagraphic sungrazers into the
expanded Kreutz group and proposed some rather unusual orbital schemes
to accommodate the evolutionary aspects of his model.  Significantly,
MacQueen \& St.\ Cyr (1991) noted that none of the SMM sungrazers
survived perihelion; the same applied to the Solwind objects as well.

With the launch of SOHO, the sungrazer research began to pick up
speed and soon reached a frantic pace.  For the understanding of the
stream of SOHO sungrazers, an important contribution at this time was
the recognition that fragmentation was not limited to perihelion but
proceeded along the entire length of the orbit (Sekanina 2000, 2002).

The next major step forward was Sekanina \& Kracht's (2015) solution
to the line-of-apsides paradox, which led to the discovery of
enormous nongravitational accelerations that the small SOHO sungrazers
were subjected to, and subsequently to a new interpretation of Marsden's
gravitational elements for these objects, including a straightforward
determination of the true longitude of the ascending node from a plot
of the nominal perihelion latitude against the nominal nodal longitude.
As an unexpected bonus, this procedure, implemented for a set of select
Kreutz sungrazers imaged exclusively in the C2 coronagraph (to curtail
scatter in the data), proved instrumental in pursuing new avenues of
research on the Kreutz system, including the discovery of its nine
populations (Sekanina 2021, 2022), trippling the number of subgroups
previously introduced by Marsden.  Each population is described by a
narrow range of true longitudes of the ascending node, consistent
with the nodal longitudes of the few bright, naked-eye sungrazers.

The discrimination of the SOHO Kreutz sungrazers into the populations
allowed me to investigate the temporal distribution of objects in each
population on the scale of years between 2000 and 2009 by constructing
bias-avoiding histograms.  Population~I showed a moderate peak in 2006
on a slowly increasing background, while Population~Ia displayed a
sharp increase throughout the decade.  In general, the histograms
suggested that the arrival rates varied substantially with time, with
intervening minima and maxima.

Strong variations in the rates of arriving SOHO sungrazers prompted
me to search for what I call swarms.  To measure a degree of scatter,
averaged over a swarm, in the true nodal longitude and
time (represented by the true perihelion time), I introduced the
{\it dispersions\/}.  Unfortunately, because as yet there
is no way to convert a nominal perihelion distance into a true perihelion
distance (by accounting, at least approximately, for the major effect of
the nongravitational acceleration), no dispersion in this element could
be introduced.

The investigation of the distribution of SOHO Kreutz
sungrazers in the true longitude of the ascending node began with
Population~I.  The histogram showed it to apparently consist of three
swarms, centered on a longitude very close to that of the Great March
Comet of 1843.  Of particular interest were the results of a search
for potential swarms in very narrow windows of the nodal longitude,
between 0$^\circ\!$.05 and 0$^\circ\!$.2; they showed that in each
of these groups of Population~I sungrazers, the dispersion in time
varied inversely as the dispersion in the nodal longitude, contrary
to expectation.  The two swarms of Population \mbox{Pre-I} displayed
the same kind of anomaly.

The examination of the SOHO Kreutz sungrazers that were arriving in
tight formation showed that the dispersion in time could be as short
as $\sim$0.25~day for swarms consisting of four members.  In general,
the dispersion in the nodal longitude was increasing with increasing
dispersion in time, so these swarms behaved differently from the ones
described above, apparently an attribute of the formation process of
the stream of SOHO Kreutz sungrazers.

The limiting magnitude of the SOHO's C2 coronagraph provides a measure
for the current lower limit to the size of boulders detected in the
stream of sungrazers.  Because of both better viewing geometry and a
more powerful instrumentation, future Sun-exploring spacecraft will
undoubtedly be able to detect much fainter Kreutz sungrazers, as
the early experience with the Parker Solar Probe appears to suggest.
A model of the stream of SOHO Kreutz sungrazers may do well to take this
prediction into account.\\[-0.25cm]

\begin{center} 
{\footnotesize REFERENCES}
\end{center}
\vspace{-0.4cm}
\begin{description}
{\footnotesize
\item[\hspace{-0.3cm}]
Battams, K., \& Knight, M.\ M.\ 2017, Phil.\ Trans.\ Roy.\ Soc.~A,~375,{\linebreak}
 {\hspace*{-0.6cm}}20160257
\\[-0.57cm]
\item[\hspace{-0.3cm}]
Ho, P.-Y.\ 1962, Vistas Astron., 5, 127
\\[-0.57cm]
\item[\hspace{-0.3cm}]
Hubbard, J.\ S.\ 1849, AJ, 1, 10
\\[-0.57cm]
\item[\hspace{-0.3cm}]
Hubbard, J.\ S.\ 1850, AJ, 1, 24, 25, 57, 153
\\[-0.57cm]
\item[\hspace{-0.3cm}]
Hubbard, J.\ S.\ 1851, AJ, 2, 46, 57
\\[-0.57cm]
\item[\hspace{-0.3cm}]
Hubbard, J.\ S.\ 1852, AJ, 2, 153
\\[-0.57cm]
\item[\hspace{-0.3cm}]
Kreutz, H.\ 1891, Publ.\ Kiel Sternw., 6
\\[-0.57cm]
\item[\hspace{-0.3cm}]
Kreutz, H.\ 1901, Astron.\ Abhandl., 1, 1
\\[-0.57cm]
\item[\hspace{-0.3cm}]
MacQueen, R.\ M., \& St.\ Cyr, O.\ C.\ 1991, Icarus, 90, 96
\\[-0.57cm]
\item[\hspace{-0.3cm}]
Marsden, B.\ G.\ 1967, AJ, 72, 1170
\\[-0.57cm]
\item[\hspace{-0.3cm}]
Marsden, B.\ G.\ 1989, AJ, 98, 2306
\\[-0.57cm]
\item[\hspace{-0.3cm}]
Marsden, B.\ G., \& Williams, G.\ V.\ 2008, Catalogue of Cometary{\linebreak}
 {\hspace*{-0.6cm}}Orbits 2008 (17th ed.)  Cambridge, MA:\ IAU Central Bureau
 for{\linebreak}
 {\hspace*{-0.6cm}}Astronomical Telegrams and Minor Planet Center, 195pp
\\[-0.57cm]
%
%
\item[\hspace{-0.3cm}]
Sekanina, Z.\ 2000, ApJ, 542, L147
\\[-0.57cm]
\item[\hspace{-0.3cm}]
Sekanina, Z.\ 2002, ApJ, 566, 577
\\[-0.57cm]
\item[\hspace{-0.3cm}]
Sekanina, Z.\ 2021, eprint arXiv:2109.01297
\\[-0.57cm]
\item[\hspace{-0.3cm}]
Sekanina, Z.\ 2022, eprint arXiv:2212.11919
\\[-0.57cm]
\item[\hspace{-0.3cm}]
Sekanina, Z., \& Chodas, P.\ W.\ 2008, ApJ, 687, 1415
\\[-0.57cm]
\item[\hspace{-0.3cm}]
Sekanina, Z., \& Chodas, P.\ W.\ 2012, ApJ, 757, 127
\\[-0.57cm]
\item[\hspace{-0.3cm}]
Sekanina, Z., \& Kracht, R.\ 2015, ApJ, 801, 135
\\[-0.625cm]
\item[\hspace{-0.3cm}]
Sekanina, Z., \& Kracht, R.\ 2022, eprint arXiv:2206.10827}
%
%
\vspace{-0.42cm}
\end{description}
\end{document}